\documentclass[pra,10pt,superscriptaddress,footnoteinbib]{revtex4}
\usepackage{amsmath}
\usepackage{latexsym}
\usepackage{amssymb}
\usepackage{graphics,epstopdf}
\usepackage[colorlinks=true, citecolor=blue, urlcolor=blue ]{hyperref}
\usepackage{epsf,graphics,graphicx}
\usepackage{float}

\newcommand{\n}{\noindent}

\newcommand{\beq}{\begin{equation}}
\newcommand{\eeq}{\end{equation}}

\begin{document}
\title{Kane model parameters and stochastic spin current}
\author{Debashree Chowdhury\footnote{Electronic address:{debashreephys@gmail.com}}} \affiliation{Physics and
Applied Mathematics Unit, Indian Statistical Institute, 203
B.T.Road, Kolkata 700108, India}

%%\vspace*{1cm}

\begin{abstract}
\n
The spin current and spin conductivity is computed through thermally driven stochastic process. By evaluating the Kramers equation and with the help of $\vec{k}.\vec{p}$ method we have studied the spin Hall scenario. Due to the thermal assistance, the Kane model parameters get modified, which consequently modulate the spin orbit coupling(SOC). This modified SOC causes the spin current to change in a finite amount.
\end{abstract}
\maketitle
%\n ~~~~~Keywords: A. Semiconductor; D.Kane model; D. Stochastic Spin current; D. Kramers equation 
\section{Introduction}
Today's condensed matter research mainly relies on the study of the spin related issues of different materials. This put forward the concept of  "Spintronics" \cite{wolf, zutic}, which unveils the importance of the spin degrees of freedom of electron for improved spin based devices. 
In this context, the development in the arena of semiconductor spintronics attracts the attention of many theoreticians as well as experimentalists. Spin Hall effect(SHE) \cite{sh1,spinh} and spin orbit coupling(SOC) are the most important candidates for many theoretical understanding in the realm of semiconductor spintronics. The spin Hall effect is the spin analogue of charge Hall effect with some differences as well. In spin Hall effect, external magnetic field is redundant for separating spin up and down spin electrons. The candidate responsible for the separation is SOC, which effectively generates a magnetic field in the rest frame of the electron and as a consequence we have spin current in this system. SOC, which is the relativistic coupling between orbital and spin degrees of freedom of electron can be obtained through the Foldy-Wouthuysen (FW) transformation \cite{m} of the Dirac equation in presence of the external electric field. Alternatively, synthetic SOC can be generated via the strain parameter or with mechanical parameters like acceleration and rotation \cite{bc,cb,cbs}. 

Besides, in semiconductor, the spin dynamics is influenced by the $\vec{ k} . \vec{ p}$ perturbation theory \cite{winkler}. The semiconductor band structure, close to band edges can be very well described by the $ \vec{ k} . \vec{ p}$ method. It is possible to explain the spin dynamics of semiconductor by taking into account the interband mixing via $ \vec{ k} . \vec{ p}$ perturbation theory. In \cite{bc}, we have demonstrated that when band structure of semiconductor is considered, the free electron SOC parameter gets modified by the Kane model parameters. The inclusion of this renormalized SOC parameter makes the theory of electron in semiconductor more accurate.

Finite temperature effects are very important issues in different aspects of spin physics \cite{sh2}. This gives the birth of spincaloritronics \cite{spinc}. Very recently, the spin Hall conductivity is demonstrated in room temperature \cite{room}, which motivates us to study the thermally activated spin related issues. We are considering here the Fokker-Planck equation for analyzing the semi-classical motion of charge carriers. We have incorporated the additional constraints like damping force and also have included stochastic force arising due to the coupling of the system with a stochastic source of heat bath.  In our formalism, the temperature correction is arising through the damping force.  Besides, in presence of temperature the Kane model parameters are affected as well. This consequently affects the spin orbit coupling and electron "g" factor. SOC is an important ingredient to have control over different physical parameters like spin current, conductivity, Berry curvature, spin relaxation time \cite{rep} e.t.c. In this paper, we theoretically have investigated the thermally driven spin current in a semiconductor on the basis of $\vec{ k} . \vec{ p}$ perturbation theory from a generalized spin orbit Hamiltonian, which includes the stochastic force and arbitrary damping force. Here the Fokker-Planck or Kramer's equation is employed to calculate the spin current. Furthermore, the effect of the crystal symmetry is also taken care of. At first, we have considered the temperature correction due to scattering mechanism through the damping constant $\gamma.$ Secondly, the temperature dependence of the Kane model parameters is appraised. Our goal is to examine the expression of the thermally assisted spin current and spin conductivity in semi-conducting system.    

The organization of the paper is as follows: in section II we build our model Hamiltonian considering the $\vec{ k} . \vec{ p}$ coupling between the $\Gamma_{6} $ conduction band and $\Gamma_{8} $ and $\Gamma_{7} $ valance bands. In section III, the semi-classical equation of motion is calculated applying the Kramer's equation. We incorporate the effect of temperature through damping constant in this section and have computed the spin current. In IV, the renormalization of the Kane model parameters through temperature is taken care of, which modifies the SOC parameters as well. This renormalization of the SOC parameter alters the spin current in a different manner than that of the previous case. The conclusion is presented in section V.

\section{The Model Hamiltonian}
The Pauli-Schr\"{o}dinger Hamiltonian with the effect of spin orbit coupling due to an external electric field can be written as \cite{bc,cb,cbs}
\beq H = \frac{\hbar^{2}\vec{k}^{2}}{2m} + qU(\vec{r}) + q\lambda\vec{\sigma} .(\vec{ k}\times \vec{ E}) + g\mu \vec{\sigma}.\vec{B} ,\label{Hk} \eeq
where the first and second terms are the kinetic term with $m$ as the free electron mass and the potential of the external electric field respectively. The third term is the spin orbit coupling term and forth term denotes the Zeeman term appearing as a consequence of external magnetic field. The free electron Hamiltonian in (\ref{Hk}) modifies significantly when we consider the whole picture within a semiconductor, where one should incorporate the $ 8\times 8 $ Kane model \cite{kane} to include the effect of energy bands. The Hamiltonian for the $ 8\times 8 $ Kane model can be written as \cite{winkler,bc}
\begin{eqnarray}
H_{8 \times 8}  =  \left( \begin{array}{ccr}
H_{6c6c} & H_{6c8v} & H_{6c7v} \\
H_{8v6c} & H_{8v8v} & H_{8v7v}\\
H_{7v6c} & H_{7v8v} & H_{7v7v}
\end{array} \right)~~~~ \\
~~=  \left( \begin{array}{ccr}
(E_{c} + eU)I_2 & \sqrt{3}P\vec{ T} . \vec{ k} & -\frac{P}{\sqrt{3}}\vec{ \sigma} . \vec{ k} \\
\sqrt{3}P \vec{ T}^{\dag} . \vec{ k} & (E_{v} + eU)I_4  & 0 \\
-\frac{P}{\sqrt{3}}\vec{ \sigma} . \vec{ k} & 0  & (E_{v} - \triangle_{0} + eU)I_{2}
\end{array} \right),
\end{eqnarray}
$\vec{T}$ matrices are given as
\begin{equation}
T_{x}  = \frac{1}{3\sqrt{2}} \left( \begin{array}{ccrr}
-\sqrt{3} & 0 & 1 & 0 \\
0 & -1 & 0 & \sqrt{3}
\end{array} \right),~~~~
T_{y}  = -\frac{i}{3\sqrt{2}} \left( \begin{array}{ccrr}
\sqrt{3} & 0 & 1 & 0 \\
0 & 1 & 0 & \sqrt{3}
\end{array} \right),
T_{z}  = \frac{\sqrt{2}}{3} \left( \begin{array}{ccrr}
0 & 1 & 0 & 0 \\
0 & 0 & 1 & 0
 \end{array}\right)
\end{equation}
and $I_{2}, I_{4}$ are unit matrices of size $2$
and $4$ respectively. $U = V_{e}(\vec{r})+ V_{c}(r),$ is the total potential of the system which contains potential due to the external electric field $V_{e}(\vec{r})$ and crystal potential $V_{c}(r)$. $E_{c}$ and $E_{v}$ denote the energies at the conduction and valence band edges respectively. $ \triangle_{0}$ is the spin orbit gap, $P$ is the Kane momentum matrix element which couples $s$ like conduction bands with $p$ like valence bands. This Kane Momentum matrix element remains almost constant for group III -- V semiconductors, whereas $ \triangle_{0}$ and $E_{G}=E_c-E_v$ varies with materials. Here $E_{G}$ denotes the energy gap between the conduction and valance band. The parameters $P$, $ \triangle_{0}$ and $E_{G}$ are known as the Kane model parameters. 
The Hamiltonian (3) can now be reduced to an effective Hamiltonian of the conduction band electron states \cite{winkler,bc} as
\beq
H_{kp} = \frac{P^2}{3}\left(\frac{2}{E_{G}} + \frac{1}{E_{G} + \triangle_{0}}\right)k^{2} + eV(\vec{r}) - \frac{P^2}{3}\left(\frac{1}{E_{G}} - \frac{1}{(E_{G} + \triangle_{0})}\right)\frac{ie}{\hbar}\vec{\sigma}.(\vec{k}\times \vec{k})\\ + e\frac{P^2}{3}\left(\frac{1}{E_{G}^{2}} - \frac{1}{(E_{G} + \triangle_{0})^{2}}\right)\vec{\sigma}.(\vec{k}\times \vec{E})\label{kp}\eeq
Now to find out the total Hamiltonian, we must add up the Hamiltonian (\ref{Hk}) with (\ref{kp}).  
The total Hamiltonian for the electron in the conduction band edges can be written as \cite{winkler}
\beq H_{tot} = \frac{\hbar^{2}\vec{k}^{2}}{2m^*} + eV(\vec{r}) + e(\lambda + \delta \lambda)\vec{ \sigma} .(\vec{ k}\times \vec{ E}) + (1+\frac{\delta g}{2})\mu \vec{\sigma}.\vec{B} ,\label{Hkp } \eeq where
$\frac{1}{m^*} = \frac{1}{m} + \frac{2P^2}{3\hbar^{2}}\left(\frac{2}{E_{G}} + \frac{1}{E_{G} + \triangle_{0}}\right)$ is the effective mass and
$\vec{ E} = -\vec{ \nabla} V_{e}(\vec{r})$ is the effective total electric field and $ \lambda = \frac{\hbar^{2}}{4m^{2}c^{2}}$ is the spin orbit coupling strength as considered in  vacuum. Furthermore, the perturbation parameters  $\delta \lambda$ and $\delta g$ are given by \cite{winkler}
\begin{eqnarray}\label{lam}
\delta \lambda &=& + \frac{P^2}{3}\left(\frac{1}{E_{G}^{2}} - \frac{1}{(E_{G} + \triangle_{0})^{2}}\right)\nonumber\\
\delta g &=& -\frac{4m}{\hbar^{2}}\frac{P^2}{3}\left(\frac{1}{E_{G}} - \frac{1}{E_{G} + \triangle_{0}}\right).
\end{eqnarray}
The $ \delta \lambda$ parameter is responsible for the renormalization of spin orbit coupling and the $\delta g$ term modifies the electron $g$ factor considerably. It is possible to show that this extra term in the electron $g$ factor can produce a shift in the ESR frequency. The Hamiltonian (\ref{Hkp }) can be rewritten neglecting the effect of Zeeman term as
\beq H_{tot} = \frac{\hbar^{2}k^{2}}{2m^*} + eU + e\lambda_{eff}\vec{\sigma} .(\vec{ k}\times \vec{ E}) ,\label{eff} \eeq
where $\lambda_{eff}=\lambda+\delta\lambda$ is the effective SOC term. The Hamiltonian in (\ref{eff}) is our system Hamiltonian, where the first term is the kinetic term, second term is the potential energy term and the third term denotes the SOC term.  The renormalization of the mass and the SOC indicates that when we consider the electron within a semiconductor, we must take care of these Kane model parameters as well.

The renormalized SOC parameter $\lambda_{eff}$ must influence the spin dynamics in of electron\cite{bc}. Our job is to find the spin current from equation (\ref{eff}).
 One can note that SOC is very important term in explaining the spin Hall effect. Here due to the interband mixing, the SOC term is changed. As a consequence the spin Hall current should modify as well. But how this SOC parameter is related to thermal corrections, is an important observation and we proceed to find this in section IV. But before that in section III we want to find out the spin current without incorporating the thermal corrections of SOC.

\section{Fokker-Planck equation and Spin current}    
 Considering Hamiltonian (\ref{eff}), it is possible to calculate the semi-classical equations of motion by evaluating $ \dot{\vec{r}}$ and $\dot{\vec{p}}$ via Heisenberg algebra. Before doing that, let us include the stochastic forces $\zeta(r,t),$ which appears as a consequences of other degrees of freedom as imperfection. One can also incorporate an arbitrary damping force $\kappa(r,p).$ Including all these forces, we can write the semi-classical equation of motion as
 \begin{eqnarray}
 \dot{\vec{r}} &=& \frac{1}{i\hbar}\left[r,H_{tot}\right] = \frac{\vec{p}}{m^{*}} + e\lambda_{eff}(\vec{\sigma}\times \vec{\nabla}U)\nonumber\\
\dot{\vec{p}} &=& -\frac{1}{i\hbar}\left[p,H_{tot}\right] = -\vec{\nabla}U - e\lambda_{eff}\nabla\left[\vec{p}.(\vec{\sigma}\times \vec{\nabla}U)\right] - \kappa(r,p) +  \zeta(t).
 \end{eqnarray}  
 As we are focusing in the linear response regime, it is suitable to consider $ \zeta$ as distributed by Gaussian white noise with an arbitrary noise strength $D.$ Thus the Fokker-Planck equation for the probability density P(r,p,t) can be obtained as \cite{Gar}
 \beq \frac{\partial P}{\partial t} = -\frac{\partial}{\partial r}\left[\frac{\partial H}{\partial p}P\right] + \frac{\partial}{\partial p}\left[\left(\frac{\partial H}{\partial r} + \kappa\right)P\right] + D\frac{\partial^2}{\partial p^2}P  .\label{kra}\eeq 
The equilibrium solution can thus be written as
\beq P(r,p) \propto exp(\frac{H}{KT}).\eeq From definition, we can write $\kappa = \frac{D}{KT}\frac{\partial H}{\partial p} = \gamma \left(\vec{p} + m^{*}\lambda_{eff}(\vec{\sigma} \times \vec{\nabla}U)\right),$ where $\gamma = \frac{D}{m^{*}KT}.$ We have to find out the steady state solutions, $K$ is the Boltzmann constant.  
Here we can write eqn. (\ref{kra}) as follows
\beq \frac{\partial P}{\partial t} = -\frac{\partial}{\partial r}(\dot{\vec{r}}P) + \frac{\partial}{\partial p}((\dot{\vec{p}} + \kappa)P) + D\frac{\partial^2}{\partial p^2}P = -\frac{\partial }{\partial r}(\frac{\pi P}{m^{*}}) -\frac{\partial}{\partial p}(AP) + D\frac{\partial^2}{\partial p^2}P,\eeq
where due to the presence of SOC, the momentum $\vec{p}$ is modified as \beq \vec{\pi} = \vec{p} + e\lambda_{eff}m^{*}(\vec{\sigma}\times \vec{\nabla}U)\eeq and $\vec{A} = -\vec{\nabla}U - e\lambda_{eff}\nabla\left[\vec{p}.(\vec{\sigma}\times \vec{\nabla}U)\right] - \vec{\kappa}(r,p).$

The time derivative of $\vec{\sigma}$ as
\beq \frac{d \vec{\sigma}}{dt} = \frac{i}{\hbar}\left[H, \sigma\right] = \frac{2}{\hbar}(\vec{\nabla}U\times \vec{p})\times \vec{\sigma}.\eeq 
Including all these quantities the force equation has the following form
\beq \langle F\rangle = \frac{d<\pi>}{dt} = \langle A\rangle + e\lambda_{eff}\langle \pi.\nabla(\vec{\sigma}\times \vec{\nabla}U) \rangle.\eeq 
\begin{figure}
\includegraphics[width=7.0 cm]{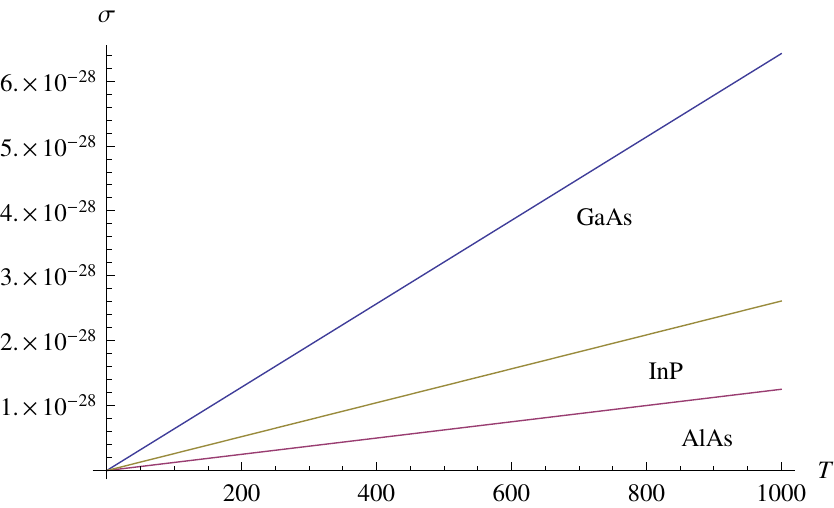}
\caption{\label{a} (Color online) Variation of the ratio of spin to charge conductivity with temperature for different semiconductors. Here $\sigma = \frac{D\sigma_{s}}{\mu K \sigma_{c}}.$ The unit in the x axis is Kelvin and in the y axis is $\mathring{A}^{-2}Kg^{-1}$}
\end{figure}
For a steady current, the average velocity of the carriers is fixed. The current in this case can be written as \beq \vec{j} = \frac{ne}{m^*}\langle\pi\rangle,\eeq with the steady state constrain as 
\beq \frac{\partial}{\partial t}\langle \pi \rangle = 0.\eeq
Here $n$ is the density of charge carriers and average is taken with the steady state solution of $P.$ The steady state solution basically depends on the electric field and also on the coupling strength($\lambda_{eff}$). The expression for current for an isotropic crystal can be written as 
\beq \vec{j} = \frac{ne^{2}}{m^*\gamma}\left[\vec{E} - \frac{1}{e}\langle \vec{\nabla} V_{c}\rangle\right] + \frac{e\lambda_{eff}ne^{2}\mu(1- \alpha)}{m^{*}\gamma^{2}}(\vec{\sigma}\times \vec{\nabla} U),\eeq where the parameter $\mu$ is given as \cite{n}
\beq \left\langle \frac{\partial^{2} V_{c}}{\partial r_{\mu}\partial r_{\nu}}\right\rangle = \mu\delta_{\mu\nu}\eeq and $\alpha$ is a dimensionless parameter given as 
\beq \left\langle \nabla V_{c}\right\rangle = \alpha e\vec{E} + \frac{e\lambda_{eff}\mu}{4m^{*2}c^{2}\gamma}(1 - \alpha)(\vec{\sigma}\times \vec{E})\eeq
Incorporating this in the expression (18) we have
\beq \vec{j} = \frac{ne^{2}(1-\alpha)}{m^{*}\gamma}\vec{E} - \frac{2\lambda_{eff}ne^{2}\mu(1- \alpha)}{m^{*}\gamma^{2}}(\vec{\sigma}\times \vec{E})\eeq 
The first term corresponds to the charge current and the second term is the spin current. Thus the thermal effect is incorporated in the current via the $\gamma$ parameter. The expression of spin current can thus be written as
\beq \vec{j}_{s} = \frac{2\lambda_{eff}ne^{2}\mu m^{*}K^{2}(1- \alpha)}{D^{2}}T^{2}(\vec{\sigma}\times \vec{E}),\label{99}\eeq
where in the last step we have incorporated the value of $\gamma= \frac{D}{m^{*}KT}$ This ensures that the spin current varies with temperature as $T^{2}.$ 
Similarly the charge current can be obtained as 
\beq \vec{j}_{c} = \frac{ne^{2}(1-\alpha)}{m^{*}\gamma}\vec{E} = \frac{Kne^{2}(1-\alpha)}{D}T \vec{E}.\label{100}\eeq
Eqn. (\ref{99}) and (\ref{100}) suggests that the spin and charge conductivities is given by
\beq \sigma_{s} = \frac{2\lambda_{eff}ne^{2}\mu m^{*}K^{2}(1- \alpha)}{D^{2}}T^{2},~~\sigma_{c} = \frac{Kne^{2}(1-\alpha)}{D}T.\label{101}\eeq 
The ratio of spin to charge conductivity is given by 
\beq \frac{\sigma_{s}}{\sigma_{c}} = \frac{2\lambda_{eff} \mu Km^{*}}{D}T.\eeq

This means that the spin to charge conductivity ratio is not constant rather depends on the temperature and increases with temperature. Also this ratio depends on the material chosen via the effective SOC parameter. Lastly one should also note that the above ratio also depends on the crystal symmetry parameter $\mu.$ For anisotropic crystals this ratio should also change via the parameters associated with the crystal anisotropy \cite{noncubic}. FIG 1 shows the variation of the ratio of spin to charge conductivity with temperature for three different semiconductors. For the plot we have used the following table as\\

\begin{center}
\begin{tabular}{|*{2}{c|}l|l|}
\hline
$E_{G}(eV)$ & $\triangle_{0}(eV)$ & $P(eV\AA)$ & $m^{*} (m_{e})$ \\
\hline
GaAs = 1.519 & 0.341 & 10.493 & 0.0665 \\ \hline
AlAs = 3.13  & 0.300 & 8.97  & 0.150  \\ \hline
InP = 1.42 & 0.110 & 8.850 & 0.0803  \\
\hline
\end{tabular}\\
\end{center}
\vspace*{.5cm}
As it is evident from the Fig. 1, that the dependence of the ratio on temperature is different for the three different semiconductors. This is one of the main results of this paper. As far as our knowledge goes, there is no experimental works which incorporate such stochastic forces to the system and investigate the effect of this on spin conductivity. Our work thus can initiate a possibility to experimentally verify the results stated here. 

\section{Thermally driven effects}
In the previous section, we have considered the effect of temperature on spin current, without incorporating the corrections due to thermal effects of SOC. In this section, our goal is to address the scenario of incorporating thermal consequences due to both stochastic force and modified Kane model parameters \cite{kane}. The temperature dependence of the band gap energy $E_{G}$ can be given as \cite{varshni}
\beq E_{G} = E_{G}(0) - \frac{aT^{2}}{T+b},\eeq  where a,b are the fitting parameters or the varshni parameters \cite{varshni}, which are different for different semiconductors. The form of the energy band gap indicated in eqn. (26) is different from that of the expression in ref. \cite{Hubner 2006}, but they will lead to the same results, which is also mentioned in \cite{Hubner 2006}.
In addition to the temperature dependence of the gap parameters, the temperature dependence of the momentum matrix element $P$ should also be considered. The Kane momentum matrix element $P,$ varies with the lattice constant $a$ as $P \approx \frac{1}{a(T)}.$ Here, the effect of phonon induced fluctuation of the interatomic spacing \cite{Hubner 2006} is not included to avoid complexities in the calculations. The temperature dependence of the lattice constant can be written by the following relation \cite{Adachi, sn}
\beq a(T) = a_{1}\left[1+ \alpha_{th}(T - 300)\right],\eeq
where $\alpha_{th}$ is the linear thermal expansion coefficient and the its value corresponds to the associated semiconductor. The values of Varshni's parameters as well as of $\alpha_{th}$ for two direct gap semiconductors can be given by \cite{Adachi,afr}\\
\begin{center}
\begin{tabular}{|*{2}{c|}l|}
\hline
$a K^{-2}$ & $b(K)$ & $\alpha_{th}(K^{-1})$ \\
\hline
GaP = 5.8 $\times $ 10$^{-4}$ & 387 & 4.65 $\times$ 10$^{-6}$  \\
\hline
InP = 4.5 $\times $ 10$^{-4}$ & 335 & 4.65 $\times$ 10$^{-6}$\\\hline
\end{tabular}\\
\end{center}

\vspace*{.5cm}
The carrier concentration is also effected due to temperature as 
\beq n = n_{c} exp(\frac{E - E_{c}}{KT}),\eeq
where $n_{c} = \frac{2}{\hbar^{3}}(2\pi m^{*}KT)^{\frac{3}{2}}.$
Here $E_{F}$ is the Fermi energy and
\beq \frac{1}{m^*} = \frac{1}{m} + \frac{2P^2}{3\hbar^{2}}\left(\frac{2}{E_{G}(T)} + \frac{1}{E_{G}(T) + \triangle_{0}}\right)\eeq is the effective mass. Although the spin orbit gap $\Delta_{0}$ may also be effected by the temperature, we are considering it as constant for simplicity.
Thus incorporating eqn. (26) and (27) in eqn. (29), we have 
\beq \frac{1}{m^*} = \frac{1}{m} + \frac{2}{3\hbar^{2}a_{1}^{2}\left[1+ \alpha_{th}(T - 300)\right]^{2}} \left(\frac{2E_{G}(0) - \frac{2aT^{2}}{T+b} + \Delta_{0}}{(E_{G}(0) - \frac{aT^{2}}{T+b} + \Delta_{0})(E_{G}(0) - \frac{aT^{2}}{T+b})}\right) = \frac{1}{m} + \frac{1}{\chi(T)},\eeq
where $\frac{1}{\chi(T)} = \frac{2}{3\hbar^{2}a_{1}^{2}\left[1+ \alpha_{th}(T - 300)\right]^{2}} \left(\frac{2E_{G}(0) - \frac{2aT^{2}}{T+b} + \Delta_{0}}{(E_{G}(0) - \frac{aT^{2}}{T+b} + \Delta_{0})(E_{G}(0) - \frac{aT^{2}}{T+b})}\right).$
Also we can write 
\begin{eqnarray} 
\delta\lambda (T) &=& \frac{1}{3a_{1}^{2}\left[1+ \alpha_{th}(T - 300)\right]^{2}}\left(\frac{1}{E_{G}^{2}(T)} - \frac{1}{(E_{G}(T) + \triangle_{0})^{2}}\right)\nonumber\\
&=& \frac{1}{3a_{1}^{2}\left[1+ \alpha_{th}(T - 300)\right]^{2}}(T+b)^{2}\left(\frac{1}{(E_{G}(0)T - bE_{G}(0) - aT^{2})^{2}} - \frac{1}{(E_{G}(0)T - bE_{G}(0) - aT^{2} + \triangle_{0})^{2}}\right)\nonumber\\
&=& \xi(T).
\end{eqnarray}
\begin{figure}
\includegraphics[width=6.0 cm]{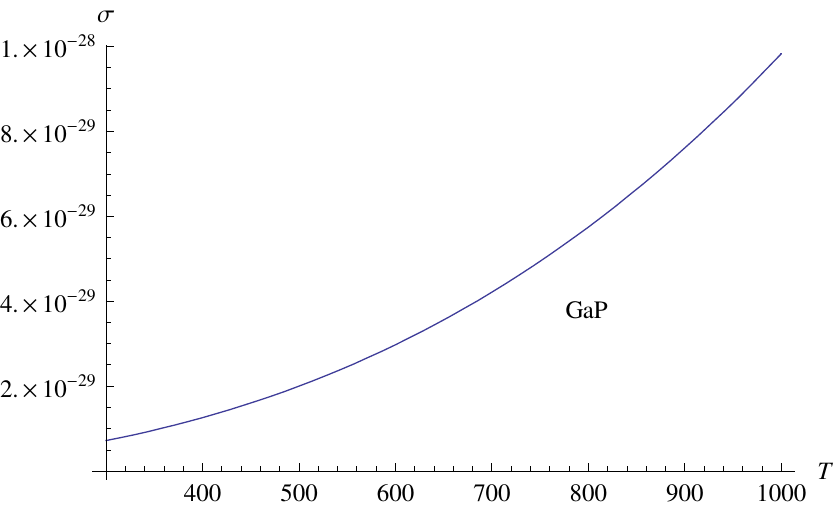}
\caption{\label{ab} (Color online) Variation of the ratio of spin to charge conductivity with temperature for GaP semiconductor, where $\sigma$ = $\frac{D\sigma_{s}}{\mu K \sigma_{c}}.$ The unit in the x axis is Kelvin and in the y axis is $\mathring{A}^{-2}Kg^{-1}$.}
\end{figure}
Thus following the same approach as in the previous section, we have the expression of spin current as 
\begin{eqnarray} \vec{j}_{s} &=& \frac{4\mu(\lambda + \xi(T))e^{2}(1-\alpha)}{\hbar\gamma^{2}}(2\pi K)^{\frac{3}{2}}(m^{*})^{\frac{5}{2}}T^{\frac{3}{2}}exp(\frac{E_{F} - E_{c}}{KT})(\vec{\sigma} \times \vec{\nabla} U)\nonumber\\
&=& \frac{4e^{2}\mu}{\hbar D^{2}}\frac{(\lambda + \xi(T))(1-\alpha)(2\pi )^{\frac{3}{2}}K^{\frac{7}{2}}}{(\frac{1}{m} + \frac{1}{\chi(T)})^{\frac{5}{2}}}T^{\frac{7}{2}}exp(\frac{E_{F} - E_{c}}{KT})(\vec{\sigma} \times \vec{\nabla} U)\nonumber\\
&=& C \frac{(\lambda + \xi(T))}{(\frac{1}{m} + \frac{1}{\chi(T)})^{\frac{5}{2}}}T^{\frac{7}{2}}exp(\frac{E_{F} - E_{c}}{KT})(\vec{\sigma} \times \vec{\nabla} U),
\end{eqnarray}
where $C = \frac{4e^{2}\mu(2\pi )^{\frac{3}{2}}K^{\frac{7}{2}}(1-\alpha)}{\hbar D^{2}}.$
The spin current obviously depends on the temperature as well as the SOC strength $\lambda$ as well.
The expression of charge current can be written as
\beq \vec{j}_{c} = \frac{AT^{\frac{5}{2}}}{(\frac{1}{m} + \frac{1}{\chi(T)})^{\frac{3}{2}}}exp(\frac{E - E_{c}}{KT})\vec{E} ,\label{110}\eeq where $A = \frac{\frac{2}{\hbar}(2\pi )^{\frac{3}{2}}K^{\frac{5}{2}}e^{2}(1-\alpha)}{D}.$
For an intrinsic semiconductor we can write the expression of spin and charge conductivity as 
\begin{eqnarray} 
\sigma_{S}(T) &=& C \frac{(\lambda + \xi(T))}{(\frac{1}{m} + \frac{1}{\chi(T)})^{\frac{5}{2}}}T^{\frac{7}{2}}exp \frac{- E_{G}(T)}{KT}\nonumber\\
\sigma_{C}(T) &=& A\frac{1}{(\frac{1}{m} + \frac{1}{\chi(T)})^{\frac{3}{2}}}T^{\frac{5}{2}}exp \frac{- E_{G}(T)}{KT}.
\end{eqnarray}
The ratio of the spin to charge conductivity can be written as 
\beq \frac{\sigma_{S}(T)}{\sigma_{C}(T)} = C_{1}\frac{(\lambda + \xi(T))}{(\frac{1}{m} + \frac{1}{\chi(T)})} T,\eeq where $C_{1} = \frac{C}{A}.$
The spin Hall conductivity varies with the temperature in a different manner. In the Previous case, it was a linear function of temperature. Figure 2 shows the variation of the dependence of the ratio of spin to charge conductivity with temperature.
This is one of our main results in this paper. In order to show the variation of this spin Hall conductivity with temperature, we can write the above equation in the following form 
\beq \sigma_{S} = C \frac{(\lambda + \xi(T))}{(\frac{1}{m} + \frac{1}{\chi(T)})^{\frac{5}{2}}}T^{\frac{7}{2}}\left(1 -\frac{E_{G}a}{4bK^{2}} - \frac{E_{G}^{2}(0)}{16bTK^{3}}-\frac{1}{T^{2}}(\frac{E_{G}^{2}}{8K^{2}}) + \frac{E_{G}^{2}}{32b^{2}K^{4}} + \frac{T^{2}a^{2}}{4b^{2}K^{2}} + \frac{aT}{2bK} - \frac{E_{G}a^{2}T}{8K^{3}b^{2}}\right).\eeq

writing the above equation only for the first order of  $E_{G}$, a and b, we have 
 
\beq \sigma_{S}(T) = C \frac{(\lambda + \xi(T))}{(\frac{1}{m} + \frac{1}{\chi(T)})^{\frac{5}{2}}}T^{\frac{7}{2}}\left(1 -\frac{E_{G}a}{4bK^{2}} + \frac{aT}{2bK}\right).\label{33}\eeq

It is quite obvious from equation (\ref{33}) that the spin conductivity depends on the temperature as well as the Varshni parameters $a,b.$ One should note that, the temperature dependence is straightforwardly arising as a result of the thermal dependence of the Kane model parameters. We are not incorporating any temperature dependence forcefully. Interestingly, we are not considering any temperature gradient to show the thermal dependence of spin current. This is the beauty of our approach.
%\subsection{Analysis of the results}

We are now in a position to analyze the results of our paper. The expression of spin conductivity in eqn. (34) can be represented as 
\begin{eqnarray} 
\sigma_{S}(T) &=& C \frac{(\lambda + \xi(T))}{(\frac{1}{m} + \frac{1}{\chi(T)})^{\frac{5}{2}}}T^{\frac{7}{2}}exp \frac{- (E_{G}(0) - \frac{aT^{2}}{T+b})}{KT}\nonumber\\
&=& C \frac{(\lambda + \xi(T))}{(\frac{1}{m} + \frac{1}{\chi(T)})^{\frac{5}{2}}}T^{\frac{7}{2}}exp\frac{- E_{G}(0)}{KT}exp\frac{aT^{2}}{KT(T+b)}.
\end{eqnarray}
One can incorporate the second exponential to the constant term $C$ and can rewrite the above expression as 
\begin{eqnarray} 
\sigma_{S}(T) &=& C^{'} \frac{(\lambda + \xi(T))}{(\frac{1}{m} + \frac{1}{\chi(T)})^{\frac{5}{2}}}T^{\frac{7}{2}}exp\frac{- E_{G}(0)}{KT},
\end{eqnarray}
where $C^{'} = Cexp\frac{aT^{2}}{KT(T+b)}.$
At high temperature the exponential term dominates over the prefactor function of temperature. So we will achieve a exponential variation of the ratio. At low temperature the term $C^{'} \frac{(\lambda + \xi(T))}{(\frac{1}{m} + \frac{1}{\chi(T)})^{\frac{5}{2}}}T^{\frac{7}{2}}$ is more dominant term. Also the result is strongly dependent of the chosen material parameters i.e on the Kane model parameters of the system. In these materials as the spin orbit gap parameter is small enough the key role is played by the energy parameter $E_{G}$ and Kane momentum matrix parameter $P,$ which have their variations with temperature.
\section{Conclusion}
In this paper, we have discussed the thermally driven effects on spin transport. Here, the $\vec{k}.\vec{p}$ perturbation theory is employed to demonstrate the spin transport issues of an electron within a semiconductor. The charge carriers are influenced by the external electric field and a Gaussian white noise. The semi-classical force equation is calculated incorporating external forces due to damping and stochastic forces. The Kramer's equation is adopted for evaluating the spin current. The temperature dependence appears due to the scattering mechanism induced in the damping factor $\gamma.$ 

It is well known that when we consider the electron transport within a semiconductor, we must take care of the band theory of the semiconductors. The inclusion of the band theory, enables us to include the Kane model parameters in the theory. In presence of temperature, the Kane model parameters are perturbed. This consequently affects the SOC parameters. The temperature dependence of the SOC parameter gives a thermally dependent spin current. The thermal dependence of spin Hall conductivity is also discussed in this paper. This shows a different approach of encountering the thermal dependence scenario in the spin transport regime.\\

\hspace{2 cm}
\begin{center}
{\bf Acknowledgment}
\end{center}
We would like to acknowledge Prof. Banasri Basu for fruitful discussions and also the anonymous referees for their valuable comments.

\end{document}